\documentclass{aa}  
\usepackage{graphicx,natbib}
\usepackage{txfonts}
\bibpunct{(}{)}{;}{a}{}{,} 

\newcommand{\kms}{\mbox{km~s$^{-1}$}}
\newcommand{\my}{\mbox{$M_{\odot}$~yr$^{-1}$}}

\begin{document}
   \title{Water vapor and silicon monoxide maser observations in the
protoplanetary nebula OH\,231.8+4.2}


   \author{J.-F. Desmurs\inst{1}
          \and
          J. Alcolea\inst{1}
          \and
          V. Bujarrabal\inst{2}
          \and
          C. S\'anchez Contreras\inst{3}
          \and
          F. Colomer \inst{2}
         }

   \offprints{J.-F. Desmurs}

   \institute{Observatorio Astron\'omico Nacional, C/Alfonso XII 3,
              E-28014 Madrid, Spain, 
              \email{jf.desmurs@oan.es}
            \and  
             Observatorio Astron\'omico Nacional, Apartado 112, Alcal\'a
              de Henares, E-28803 Madrid, Spain 
            \and
             Dpto. Astrof\'{\i}sica Molecular e Infrarroja (DAMIR),
              Instituto de Estructura de la Materia - CSIC, C/Serrano
              121,E-28006 Madrid, Spain 
             }

   \date{Received: 12/12/2006; Accepted: 09/03/2007}

 
  \abstract 
{OH\,231.8+4.2 is a well studied preplanetary nebula (pPN) around a 
binary stellar system that shows a remarkable bipolar outflow.}
{To study the structure and kinematics of the inner 10-80\,AU
nebular regions probed by SiO and H$_2$O maser emission, where the
agents of wind collimation are expected to operate, in order to gain
insights into the, yet poorly known, processes responsible for the
shaping of bipolar pPNe.}
{We performed high-resolution observations of the H$_2$O
6$_{1,6}$--5$_{2,3}$ and $^{28}$SiO $v$=2, $J$=1--0 maser emissions
with the Very Long Baseline Array. The absolute position of both
emission distributions were recovered using the phase referencing
technique, and accurately registered in HST optical images.}
{Maps of both masers were produced and compared. 
H$_2$O maser clumps are found to be distributed in two areas of 20 mas
in size spatially displaced by $\sim$60 milli-arcseconds along an axis
oriented nearly north-south.
SiO masers are tentatively found to be placed between the two H$_2$O
maser emitting regions, probably indicating the position of the
Mira component of the system.}
{The SiO maser emission traces an inner equatorial component with a
diameter of 12\,AU, probably a disk rotating around the M-type
star. Outwards, we detect in the H$_2$O data a pair of polar caps,
separated by 80\,AU. We believe that the inner regions of the nebula
probably have been altered by the presence of the companion, leading to
an equator-to-pole density contrast that may explain the lack of H$_2$O
masers and strong SiO maser emission in the denser, equatorial
regions.}


   \maketitle
%

\section{Introduction}
\label{intro}
OH\,231.8+4.2 (hereafter OH\,231.8) is a well studied pre-planetary
nebula (pPN). The central source is a binary system formed by an M9-10
III Mira variable (i.e. an AGB star) and a A0 main sequence companion,
as revealed by optical spectroscopy \citep[see][]{sanchez04}. This
remarkable bipolar nebula shows all the signs of post-AGB evolution:
fast bipolar outflows with velocities $\sim$200--400~km\,s$^{-1}$,
shock-excited gas and shock-induced chemistry. The distance of this
source, $\sim$ 1500 pc, and the inclination of the bipolar axis with
respect to the plane of the sky, $\sim 36\degr$, are well known, in
particular thanks to measurements of phase lags between the variability
of the radiation coming from the two lobes and of the light
polarization in them \citep[see][]{bowers84, kastner92, shure95}.  The
presence of a late-type star in the core of a bipolar post-AGB nebula
like OH\,231.8 is very unusual since the central stars of pPNe are
typically hotter, with spectral types from B to K.
PNe and pPNe usually present conspicuous departures from spherical
symmetry, including e.g. multiple lobes and jets. To explain their
evolution from spherical AGB envelopes, several models have postulated
the presence of dense rings or disks close to the central post-AGB
stars \citep[see][e.g.]{soker02,frankb04}. The accretion of material
from them would drive the post-AGB jets by magnetrocentrifugal
launching, in a similar way as in protostars. The interaction of
these jets with the AGB envelope give rise to shocks that shape and
accelerate the pPN lobes. Existing observations reveal the presence
of central disks in a few pPNe, but their limited spatial resolution
cannot unveil the very inner regions of the disks that are relevant for
the processes mentioned above.

From mid-infrared MIDI observations of OH\,231.8, \citet{matsuura06}
show the presence of a compact circumstellar region with an inner
radius of 40-50 AU ($\sim$35 milliarcsec (mas) at 1.3 kpc).
An equatorial torus is observed at distances larger than
1 arcsec, however, no trace of rotation is found at this scale and the
gas is known to be in expansion, as shown by CO and OH emission data
\citep[e.g.][]{alcolea01,zijlstra01}.

\begin{figure*}
  \centering
  \includegraphics[width=18cm]{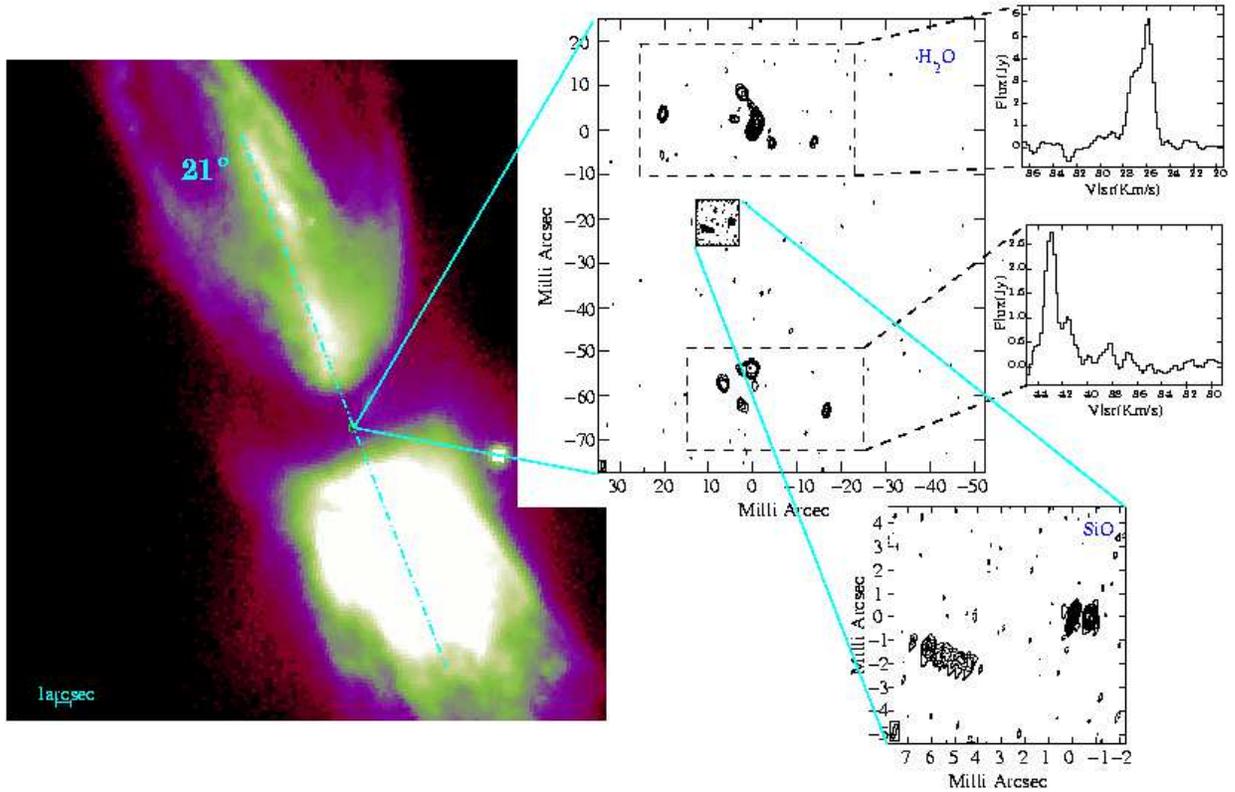}
    \caption{Total intensity map of the H$_2$O maser, compared with the
             $HST$ image of the nebula (taken with the WFPC2, filter
             F791W); accurate astrometry has been performed for the HST
             image and maser maps. The small square map indicates the
             absolute position of the SiO maser, and the SiO map
             reproduces the SiO data by \citet{sanchez02}.  Top right
             panels show the composed profile of H$_2$O maser for the
             two main regions.}
    \label{mapabsh2o}
\end{figure*}

Due to the late spectral type ($T_{\rm eff} \sim$~2500~K) of its AGB
central star, OH 231.8 still shows intense SiO masers, contrarily to
what happens in the majority of pPNe.  NRAO\footnote{The National Radio
Astronomy Observatory is a facility of the National Science Foundation
operated under cooperative agreement by Associated Universities, Inc.}
Very Long Baseline Array (VLBA) observations of $^{28}$SiO masers in
OH\,231.8, carried out at 7~mm ($v$=2, $J$=1--0) in April 2000
revealed for the first time the structure and kinematics of the close
stellar environment in a pPN \citep{sanchez02}. The SiO maser emission
arises from several compact, bright spots forming a structure elongated
in the direction perpendicular to the symmetry axis of the nebula. Such
a distribution is consistent with an equatorial torus with a radius of
$\sim$\,6~AU around the central star. A complex velocity gradient was
found along the torus, which suggests rotation and infall of material
towards the star. 
Such a distribution is remarkably different from that typically found
in maps of AGB stars, where the masers form a roughly spherical
ring-like chain of spots resulting from tangential maser amplification
in a thin, spherical shell \citep{desmurs00,cotton04}.

Water vapor masers are very often observed in AGB envelopes. They occur
in regions extending between 30 and 70 AU, therefore, about 5 times
larger than the SiO maser shells.  Although, in pPNe, water masers are
rarely observed, they have been detected in OH231.8. VLA maps of the
H$_2$O emission by \citet{gomez01} show it to be coincident in position
with the SiO masers, but their resolution (about 1 arcsec) was not
enough to resolve its structure, as expected. Here, we present VLBA
observations of these masers as well as new maps of the SiO maser
emission. In both cases we have been able to obtain the absolute
position of the spots with great accuracy, and locate them with respect
to the other (more extended) components of the nebulae.


\section{Observations and astrometry} 

             
\begin{table}
\begin{minipage}[t]{\columnwidth}
\caption{Detected H$_2$O maser components.}
\label{table1}      
\centering          
\renewcommand{\footnoterule}{}  
\begin{tabular}{c | c | c | c | c | c }     
\hline 
Num  & $\Delta$R.A.  & $\Delta$Dec & $V_{lsr}$ & Peak flux & Width \\
     &(mas)& (mas) & km~s$^{-1}$& Jy & km~s$^{-1}$\\
\hline       
1 &   0    &  0    & 25.5 & 12.3& 2.0 \\
2 &  -0.89 &  2.55 & 27.2 & 5   & 2.0 \\
3 &  20.31 &  3.72 & 29.1 & 2.5 & 1.4 \\
4 &   2.38 &  8.35 & 27.8 & 1.8 & 1.9 \\
5 &  -4.25 & -2.82 & 26.6 & 0.7 & 0.9 \\
6 & -13.86 & -2.62 & 27.4 & 0.56& 0.6 \\
7 &   4.16 & 2.46  & 25.4 & 0.4 & 0.4 \\
8 &   8    & 6     & 20.8 & 0.3 & 0.4 \\
9 &  21    & -5.0  & 30.3 & 0.2 & 0.3 \\
\hline                    
10 &   0.33 & -53.90 & 42.9 & 2.9 & 2.3 \\
11 & -16.68 & -63.36 & 38.5 & 1.5 & 2.4 \\
12 &   6.57 & -57.48 & 41.9 & 1.3 & 2.3 \\
13 &   2.55 & -62.06 & 41.6 & 0.85& 0.9 \\
14 &   3.03 & -53.84 & 43.1 & 0.59& 0.8 \\
15 &   0.1  & -57.60 & 44.4 & 1.3 & 0.6 \\
16\footnote{Outside of the map}& -58 & -54 & 40.8 & 0.1 & 0.6 \\
\hline                  
\end{tabular}
\end{minipage}
\end{table}

Using the VLBA, we performed milliarcsecond resolution observations of
the H$_2$O 6$_{1,6}$--5$_{2,3}$ and SiO ($v$=1 and $v$=2, $J$=1--0)
maser emission in the pPN OH\,231.8. The H$_2$O and SiO observations
were performed on 24th November 2002 and on 3rd July 2003,
respectively. The data were recorded in dual circular polarization with
a velocity resolution (i.e. channel width) of $\sim$0.2~km\,s$^{-1}$
and a total velocity coverage of about 110~km\,s$^{-1}$. To be able to
measure the relative position of the H$_2$O and SiO maser spots /
clumps part of the observations was made using the phase referencing
technique: to ensure the detection of our source, we alternate standard
line observations with observations in phase referencing, each
observing block lasting 40 minutes. The coordinates are then absolute
positions referred to the quasar J0730--11.  For the correlation of all
the observations, we used the coordinates given by
\citet{sanchez00a}, derived from IRAM plateau de Bure observations
(R.A.$_{\rm J2000}=07^h42^m16\fs93$ Dec$_{\rm
J2000}=-14\degr42\arcmin50\farcs2$).

The data reduction followed the standard scheme for spectral line
calibration data in the Astronomical Image Processing System
(AIPS). The amplitude was calibrated using the template spectra method.
To improve the phase calibration solutions, we extensively mapped all the
calibrators, and their brightness distributions were introduced and
taken into account in the calibration process.

\subsection{Relative position of H$_2$O and SiO emission}

In order to compare the SiO and the H$_2$O maser emission, the
observations were performed using the phase referencing technique. This
means that the antennas were rapidly switched between our source
(OH231.8) and our calibrator J0730--11 in less time than the phase
coherence time of the atmosphere at the given frequency (less than 1
minute for the full switching cycle). Therefore during the data
reduction process, we were able to directly apply the phase correction
derived from our phase calibrator onto OH231.8. (J0730--11 was well
detected at both frequencies with a flux density measured in the
resulting maps of 1.44~Jy at 22~GHz, and 1.33~Jy at 43~GHz).

For the water maser observations the phase referencing worked perfectly
and the most intense channel of the autocorrelation spectra (the peak
lying at 25.5 ~\kms\ LSR) was easily detected and mapped with a
signal-to-noise ratio over 10\,$\sigma$. The measured absolute position
lies at R.A. and Dec offsets $\Delta$(R.A.)=$-149$~mas and
$\Delta$(Dec.)=$+166$~mas from the correlation position. Therefore, the
position measured for our reference H$_2$O maser spot is
R.A.$_{\rm J2000}=07^h42^m16\fs91974$
Dec$_{\rm J2000}=-14\degr42\arcmin50\farcs0354$.

The SiO maser line ($v$=2, $J$=1--0) was weak at the time of the
observations, only a few Jy, and the phase referenced map is noisy.
The SiO maser profile in our data is composed of three peaks, at about
26, 32 and 40 \kms~LSR, similar to those observed in other epochs in
the source, see e.g.\ \citet{sanchez00a,sanchez02}.  We were able to
map the emission in the most intense channel, the one lying close to
27~\kms~LSR, and identified a feature with a flux of 72.55~mJy i.e. at
a level of 4.4~$\sigma$ detection.  This tentative detection for the
SiO maser emission give us an absolute position at offsets
$\Delta$(R.A.)=$-145$~mas and $\Delta$(Dec.)=$+145$~mas with respect to
the correlation position. The absolute coordinates of our reference SiO
feature would be: R.A.$=07^h42^m16\fs91994$
Dec$=-14\degr42\arcmin50\farcs0574$.  This would place the SiO maser
emission nearly in the middle of the two groups of water maser spots
(see Fig. \ref {mapabsh2o}). The SiO map shown in Fig.  \ref{mapabsh2o}
is that obtained by \citet{sanchez02},
aligned under the assumption that the position derived by us coincides
with the centroid of their spot distribution.  Because of the small
size of the SiO maser distribution, the uncertainty of this alignment
is in any case very small, less than about 3 mas.

In order to locate the SiO and H$_2$O maser spots in the HST image of
the nebula, we improved the astrometry of the HST images published by
some of us \citep{Bujarrabal02}. In that paper, the registering of the
images was done with IRAF using the ``STSDAS'' task ``metric'' by
comparing the positions of 28 stars in the FWPC2 chips listed in the
Guide Star Catalog 2.2 with those retrieved by the task. We concluded
that, after corrections, the positions of the plates where accurate up
to 0.2 arcsec. Since then, the Second U.S. Naval Observatory CCD
Astrograph Catalog \citep{zacharias04} has been released. We have 9
stars in the 4 FWPC2 chips listed in this catalog with position errors
of about 0.02 arcsec. or less. We have compared these nominal positions
with those derived with the task ``metric'' obtaining mean errors
(measured -- nominal) of $-0\farcs 81$ to the east ($-0.056$ sec. of
time in R.A.) and $-0\farcs 43$ in Dec, with rms of $0\farcs12$ and
$0\farcs054$ respectively \citep[fully consistent with the values of
$-0\farcs6 \pm 0\farcs2$ and $-0\farcs5 \pm 0\farcs2$ obtained
by][]{Bujarrabal02}. Using these corrections we have ploted the
location of both maser emissions on the HST plate shown in
Fig. \ref{mapabsh2o}.

\section{Results and discussion}

In Table \ref{table1} we summarise our H$_2$O observational
results. Column 2 and 3 give the offset in mas of each maser spot
respectively to the reference maser spot. This reference maser was used
to derive the fringe rate corrections, and is the one for which we
measured the absolute position with respect to the quasar. The three
last columns give the values obtained from Gauss fit of
each maser spot, central velocity, peak flux intensities, and the full
linewidth at half peak intensity.

The H$_2$O maser integrated flux map is shown in Fig. \ref{mapabsh2o}.
The restored clean beam has a full width at half maximum of about
$\sim$ 1.5 x 0.75 mas, and is oriented with a position angle of PA =
$-5\degr$.  Maser emission has been searched for a large area of up to
900 mas. Maps have been cleaned down to a noise level of 30~mJy/beam.
Maser emission arises from several compact spots distributed in 
mainly two areas, separated by about $60$~mas, in approximately the
north-south direction, corresponding to the two spectral features
appearing on the auto and cross correlation spectra (note that we also
found an isolated maser spot at a few tenths of mas to the west). The
northern area correspond to the blue-shifted emission and the southern
to the red-shifted one. Our SiO observations indicate that the SiO
masers arise from a region placed approximately in the middle of both
H$_2$O maser-emitting areas (see Fig. \ref{mapabsh2o}), probably
pinpointing the location of the central Mira component of the central
stellar system.

The size of the water maser regions around AGB stars varies between
$\sim$ 5 10$^{14}$ and 2 10$^{15}$~cm
\citep[e.g.][]{colomer00,bains03}, occupying the circumstellar region
in which the expansion velocity is already high but probably has not
yet reached its final value (around 8--10 \kms). The distribution of
the maser spots in some objects (IRC\,+60169, U Her, IK Tau, RT Vir,
...), is rounded and consistent with the maser emission arising in a
thick expanding shell. In other objects, the distribution of the maser
spot is more chaotic.

It is thought that, at a distance of a few 10$^{14}$ cm from the AGB
star, the temperature (500 -- 1000 K) and densities (5 10$^7$ -- 10$^9$
cm$^{-3}$, for standard mass-loss rates between $\sim$ 10$^{-6}$ and
10$^{-5}$ \my) are adequate to produce strong H$_2$O maser emission
\citep[see model predictions by e.g.][]{cooke85}. Water abundances in
these layers are still high, more or less equal to those given by
thermal equilibrium.

As we have seen, the size of the total emitting region in OH\,231.8,
$\sim$ 10$^{15}$ cm, is very compatible with that in standard red
giants. Also the spot relative velocity, requiring a deprojected
expansion velocity of 10 - 15 \kms\ (for an axis inclination of about
36$\degr$ with respect to the plane of the sky, Sect.~\ref{intro}), is
compatible with the usual velocities found in AGB stars
\citep[e.g.][]{loup93,telinkel89}. However, we note that the total
velocity dispersion in our data ($\sim$ 20~\kms) is larger than those
found in the majority of H$_2$O profiles from AGB stars
\citep[see e.g.][]{engels96,engels02,colomer00,bains03}.

Therefore, from the point of view of the pumping conditions and
kinematics, the H$_2$O emission in our object comes from a region that
we can assume to be typical of water masers in AGB stars. No special
phenomena (shocks, photo induced chemistry, ...) are then necesarily
required to explain our observations, just the usual physical and
chemical conditions in the envelope of a O-rich AGB star.
 
The bipolar spatial distribution of H$_2$O masers in OH\,231.8 is,
however, peculiar. Instead of a circular distribution around the central
star, we find two spot sets. The fact that the SiO maser emission comes
from a region in the middle of the H$_2$O spot areas confirms that they
are placed at both sides of the star, along the axis of the nebula.
This distribution strongly suggests that the equatorial regions of the
inner circumstellar envelope of OH\,231.8 have been strongly modified,
probably by the presence of a companion,
and are now not able to harbor water vapor masers. Running simple
modeling of water maser excitation \citep[see][]{soria04}, it appears
that water masers can be easily quenched: a modification by a factor of
2 of the radius of the emission (this corresponds to a factor 4 in
density) produces intensity drop of the maser by one order of
magnitude.

Based on its broad OH maser emission profile, OH231.8 has been
suggested to be a "water-fountain" cousin \citep[e.g.][]{likkel92}.
"Water-fountain" sources are a special class of rare post-AGB stars
with very high velocity outflows traced by OH or H$_2$O masers or
both. So far there are only four objects in this group, namely:
IRAS\,16342-3814, IRAS\,19134+2131, W43A \citep[e.g.][]{likkel92}, and
the most recently discovered OH\,12.8-0.9 \citep{boboltz05}. The
expansion velocities of the outflows implied by the H$_2$O maser
emission in these objects (except for OH\,12.8 - see below),
$V_{\mathrm{exp}}$ = 100 -- 300 \kms, are one order of magnitude larger
than the typical shell expansion velocity of mass-losing evolved stars
(see above). Another characteristic of water fountain sources is that
the expansion velocity derived from H$_2$O is higher than that deduced
from the OH maser emission. This implies that the location of the
masers in these objects deviates from the prototypical maser
stratification in AGB envelopes, where the SiO, H$_2$O, and OH masers
arise in different shell layers located at progressively further
distances from the star, expanding also at progressively higher
velocities. As revealed by recent high-angular resolution observations,
the spatial structure of the H$_2$O masers in water-fountain sources is
indeed peculiar: the masers are occur in collimated winds or jets,
aligned with the symmetry axis of the nebulae
\citep[e.g.][]{imai04,boboltz05}. These maser spots are located at
distances that range between 870~AU and 600~AU from the central star,
i.e. much farther away than H$_2$O masers in OH/IR stars (typically at
40 - 70 AU) and therefore, shock-compression and heating is required to
explain the fast maser spots at these distances: the maser emission
would be located in an area of shocked gas, in the region where the
post-AGB jets interact with the fossil AGB envelope.

Although the nebular geometry model (equatorial disk + polar stream)
proposed to explain water-fountain sources probably applies to OH231.8
to some extent, the H$_2$O maser emission in OH231.8 is remarkably
different from that found in water-fountain objects (see above).  A few
other pPNe show H$_2$O emission with low-velocity spots, more similar
to that observed in our source \citep[][]{gregorio04}. Although the
interpretation of the origin of the H$_2$O emission in them is not
straightforward, it seems to come from nebular regions that have not
been accelerated during the post-AGB evolution. In particular, the
expansion velocity of the H$_2$O masers is very low compared to pPN
bipolar jets. Moreover, the H$_2$O emission is not as well collimated
as in water-fountains.  As a consequence, shocks effects (expected to
be responsible for the pPNe bipolar outflows) are not required to
explain the H$_2$O maser emission in OH231.8, which is observed at only
$\sim$ 40 AU from the central star, as in normal AGB envelopes, and
expanding at low velocity, lower than that of the OH maser spots.

 We cannot rule out that the H$_2$O maser spots of OH231.8 are placed
at the base of a bipolar outflow, whose very fast, extended components
are detected in scattered light and CO line emission. In this case, the
masers would lie there because a bipolar ejection of material has
yielded two opposite condensations with suitable conditions to form
H$_2$O masers. But, as discussed above, the properties of water vapor
masers in our source are very different from those of maser associated
with bipolar outflows in pPNe. In fact, they resemble those of H$_2$O
masers associated with AGB stars.  We also note that if the H$_2$O
masers are tracing a bipolar outflow, this would be expected to be
centered around the compact companion of the binary system, which is
most likely to power the jets in this case \citep[e.g.][]{sanchez04},
and not around the mass-losing star, i.e. the Mira, as we observe.

In summary, in view of their spatial distribution and kinematics, the
maser spots in OH\,231.8 seem to be the result of a pumping process
very similar to that at work in AGB circumstellar envelopes, and under
similar physical and chemical conditions. But they take place in an
envelope that has been seriously altered, probably due to the presence
of a companion, in such a way that only (unshocked) polar regions of
the envelope keep favorable conditions for H$_2$O maser pumping. The
fact that the SiO maser region is much smaller than the H$_2$O region
and perpendicular to its distribution, probably indicates that
gravitational/tidal forces due to the companion strongly affect these
inner circumstellar shells, supporting our conclusion on its effects on
the H$_2$O emitting region.

\bibliographystyle{aa}
\bibliography{references}

\begin{thebibliography}{28}
\expandafter\ifx\csname natexlab\endcsname\relax\def\natexlab#1{#1}\fi

\bibitem[{{Alcolea} {et~al.}(2001){Alcolea}, {Bujarrabal}, {S{\'a}nchez
  Contreras}, {Neri}, \& {Zweigle}}]{alcolea01}
{Alcolea}, J., {Bujarrabal}, V., {S{\'a}nchez Contreras}, C., {Neri}, R., \&
  {Zweigle}, J. 2001, \aap, 373, 932

\bibitem[{{Bains} {et~al.}(2003){Bains}, {Cohen}, {Louridas}, {Richards},
  {Rosa-Gonz{\'a}lez}, \& {Yates}}]{bains03}
{Bains}, I., {Cohen}, R.~J., {Louridas}, A., {et~al.} 2003, \mnras, 342, 8

\bibitem[{{Boboltz} \& {Marvel}(2005)}]{boboltz05}
{Boboltz}, D.~A. \& {Marvel}, K.~B. 2005, \apjl, 627, L45

\bibitem[{{Bowers} \& {Morris}(1984)}]{bowers84}
{Bowers}, P.~F. \& {Morris}, M. 1984, \apj, 276, 646

\bibitem[{{Bujarrabal} {et~al.}(2002){Bujarrabal}, {Alcolea}, {S{\'a}nchez
  Contreras}, \& {Sahai}}]{Bujarrabal02}
{Bujarrabal}, V., {Alcolea}, J., {S{\'a}nchez Contreras}, C., \& {Sahai}, R.
  2002, \aap, 389, 271

\bibitem[{{Colomer} {et~al.}(2000){Colomer}, {Reid}, {Menten}, \&
  {Bujarrabal}}]{colomer00}
{Colomer}, F., {Reid}, M.~J., {Menten}, K.~M., \& {Bujarrabal}, V. 2000, \aap,
  355, 979

\bibitem[{{Cooke} \& {Elitzur}(1985)}]{cooke85}
{Cooke}, B. \& {Elitzur}, M. 1985, \apj, 295, 175

\bibitem[{{Cotton} {et~al.}(2004){Cotton}, {Mennesson}, {Diamond}, {Perrin},
  {Coud{\'e} du Foresto}, {Chagnon}, {van Langevelde}, {Ridgway}, {Waters},
  {Vlemmings}, {Morel}, {Traub}, {Carleton}, \& {Lacasse}}]{cotton04}
{Cotton}, W.~D., {Mennesson}, B., {Diamond}, P.~J., {et~al.} 2004, \aap, 414,
  275

\bibitem[{{de Gregorio-Monsalvo} {et~al.}(2004){de Gregorio-Monsalvo},
  {G{\'o}mez}, {Anglada}, {Cesaroni}, {Miranda}, {G{\'o}mez}, \&
  {Torrelles}}]{gregorio04}
{de Gregorio-Monsalvo}, I., {G{\'o}mez}, Y., {Anglada}, G., {et~al.} 2004,
  \apj, 601, 921

\bibitem[{{Desmurs} {et~al.}(2000){Desmurs}, {Bujarrabal}, {Colomer}, \&
  {Alcolea}}]{desmurs00}
{Desmurs}, J.~F., {Bujarrabal}, V., {Colomer}, F., \& {Alcolea}, J. 2000, \aap,
  360, 189

\bibitem[{{Engels}(2002)}]{engels02}
{Engels}, D. 2002, \aap, 388, 252

\bibitem[{{Engels} \& {Lewis}(1996)}]{engels96}
{Engels}, D. \& {Lewis}, B.~M. 1996, \aaps, 116, 117

\bibitem[{{Frank} \& {Blackman}(2004)}]{frankb04}
{Frank}, A. \& {Blackman}, E.~G. 2004, \apj, 614, 737

\bibitem[{{G{\'o}mez} \& {Rodr{\'{\i}}guez}(2001)}]{gomez01}
{G{\'o}mez}, Y. \& {Rodr{\'{\i}}guez}, L.~F. 2001, \apjl, 557, L109

\bibitem[{{Imai} {et~al.}(2004){Imai}, {Morris}, {Sahai}, {Hachisuka}, \&
  {Azzollini F.}}]{imai04}
{Imai}, H., {Morris}, M., {Sahai}, R., {Hachisuka}, K., \& {Azzollini F.},
  J.~R. 2004, \aap, 420, 265

\bibitem[{{Kastner} {et~al.}(1992){Kastner}, {Weintraub}, {Zuckerman},
  {Becklin}, {McLean}, \& {Gatley}}]{kastner92}
{Kastner}, J.~H., {Weintraub}, D.~A., {Zuckerman}, B., {et~al.} 1992, \apj,
  398, 552

\bibitem[{{Likkel} {et~al.}(1992){Likkel}, {Morris}, \& {Maddalena}}]{likkel92}
{Likkel}, L., {Morris}, M., \& {Maddalena}, R.~J. 1992, \aap, 256, 581

\bibitem[{{Loup} {et~al.}(1993){Loup}, {Forveille}, {Omont}, \&
  {Paul}}]{loup93}
{Loup}, C., {Forveille}, T., {Omont}, A., \& {Paul}, J.~F. 1993, \aaps, 99, 291

\bibitem[{{Matsuura} {et~al.}(2006){Matsuura}, {Chesneau}, {Zijlstra}, {Jaffe},
  {Waters}, {Yates}, {Lagadec}, {Gledhill}, {Etoka}, \&
  {Richards}}]{matsuura06}
{Matsuura}, M., {Chesneau}, O., {Zijlstra}, A.~A., {et~al.} 2006, \apjl, 646,
  L123

\bibitem[{{S{\'a}nchez Contreras} {et~al.}(2000){S{\'a}nchez Contreras},
  {Bujarrabal}, {Neri}, \& {Alcolea}}]{sanchez00a}
{S{\'a}nchez Contreras}, C., {Bujarrabal}, V., {Neri}, R., \& {Alcolea}, J.
  2000, \aap, 357, 651

\bibitem[{{S{\'a}nchez Contreras} {et~al.}(2002){S{\'a}nchez Contreras},
  {Desmurs}, {Bujarrabal}, {Alcolea}, \& {Colomer}}]{sanchez02}
{S{\'a}nchez Contreras}, C., {Desmurs}, J.~F., {Bujarrabal}, V., {Alcolea}, J.,
  \& {Colomer}, F. 2002, \aap, 385, L1

\bibitem[{{S{\'a}nchez Contreras} {et~al.}(2004){S{\'a}nchez Contreras}, {Gil
  de Paz}, \& {Sahai}}]{sanchez04}
{S{\'a}nchez Contreras}, C., {Gil de Paz}, A., \& {Sahai}, R. 2004, \apj, 616,
  519

\bibitem[{{Shure} {et~al.}(1995){Shure}, {Sellgren}, {Jones}, \&
  {Klebe}}]{shure95}
{Shure}, M., {Sellgren}, K., {Jones}, T.~J., \& {Klebe}, D. 1995, \aj, 109, 721

\bibitem[{{Soker}(2002)}]{soker02}
{Soker}, N. 2002, \apj, 568, 726

\bibitem[{{Soria-Ruiz} {et~al.}(2004){Soria-Ruiz}, {Alcolea}, {Colomer},
  {Bujarrabal}, {Desmurs}, {Marvel}, \& {Diamond}}]{soria04}
{Soria-Ruiz}, R., {Alcolea}, J., {Colomer}, F., {et~al.} 2004, \aap, 426, 131

\bibitem[{{Te Lintel Hekkert} {et~al.}(1989){Te Lintel Hekkert},
  {Versteege-Hensel}, {Habing}, \& {Wiertz}}]{telinkel89}
{Te Lintel Hekkert}, P., {Versteege-Hensel}, H.~A., {Habing}, H.~J., \&
  {Wiertz}, M. 1989, \aaps, 78, 399

\bibitem[{{Zacharias} {et~al.}(2004){Zacharias}, {Urban}, {Zacharias},
  {Wycoff}, {Hall}, {Monet}, \& {Rafferty}}]{zacharias04}
{Zacharias}, N., {Urban}, S.~E., {Zacharias}, M.~I., {et~al.} 2004, \aj, 127,
  3043

\bibitem[{{Zijlstra} {et~al.}(2001){Zijlstra}, {Chapman}, {te Lintel Hekkert},
  {Likkel}, {Comeron}, {Norris}, {Molster}, \& {Cohen}}]{zijlstra01}
{Zijlstra}, A.~A., {Chapman}, J.~M., {te Lintel Hekkert}, P., {et~al.} 2001,
  \mnras, 322, 280

\end{thebibliography}
\end{document}